\documentclass[5p]{elsarticle}
\usepackage{mathptmx} % assumes new font selection scheme installed
\usepackage{times} % assumes new font selection scheme installed
\usepackage{amsmath} % assumes amsmath package installed
\usepackage{amssymb}  % assumes amsmath package installed
\usepackage{psfrag}
\usepackage{pmat}
\usepackage[table]{xcolor}
\usepackage{algorithmic}
\usepackage{array}
\usepackage{stfloats}
\usepackage{graphics} % for pdf, bitmapped graphics files
\usepackage{graphicx}
\usepackage{subfigure} % locat two figures into one picture, and especially command \figure*
\usepackage{algorithm}% for the algorithm
\usepackage{epsfig} % for postscript graphics files
\usepackage{mathptmx} % assumes new font selection scheme installed
\usepackage{times} % assumes new font selection scheme installed
\usepackage{amsmath} % assumes amsmath package installed
\usepackage{amssymb}  % assumes amsmath package installed
\usepackage{float} %用于固定图片位置
\usepackage{footnote}
\makesavenoteenv{tabular}
\usepackage{threeparttable}

\newtheorem{theorem}{Theorem}
\newtheorem{corollary}[theorem]{Corollary}

\newtheorem{remark}{Remark}
\newproof{pf}{Proof}

\usepackage{verbatim,pifont,color,amssymb,amsfonts,amsmath,graphicx,pgf,slashbox,psfrag,ifpdf}
\usepackage[mathscr]{eucal}

\def\eg{{\it e.g., \/}}
\def\defeq{\triangleq}
\def\T{\mbox{\tiny T}}
\def\N{\mbox{\tiny N}}
\def\K{\mbox{\tiny K}}

\title{A Subspace Technique for The Identification of Switched Affine Models}
\author[auto,info]{Liang Li} \ead{liang-li07@mails.tsinghua.edu.cn}
\author[info]{Wei Dong}                   \ead{weidong@mail.tsinghua.edu.cn}
\author[info]{Yindong Ji}                 \ead{jyd@mail.tsinghua.edu.cn}
\author[corn]{Lang Tong\corref{cor1}}                  \ead{ltong@ece.cornell.edu}

\cortext[cor1]{Corresponding author, email:ltong@ece.cornell.edu, fax: (607)255-9072. This paper was not presented at any IFAC meeting.}
\address[auto]{Department of Automation, Tsinghua University, Beijing 100084, China}
\address[info]{National Laboratory for Information Science and Technology, Tsinghua University, Beijing 100084, China}
\address[corn]{School of Electrical and Computer Engineering, Cornell University, Ithaca, NY 14853, USA}

\begin{document}

\begin{frontmatter}

%%%%%%%%%%%%%%%%%%%%%%%%%%%%%%%%%%%%%%%%%%%%%%%%%%%%%%%%%%%%%%%%%%%%%%%%%%%%%%%%

\begin{abstract}
The problem of estimating parameters of switched affine systems with noisy input-output observations is considered.
The switched affine models is transformed into a switched linear one by removing its intersection subspace, which is estimated from observations.
A subspace technique is proposed to exploit the observations' permutation structure, which transforms the problem of associating observations with subsystems into one of de-permutating a block diagonal matrix, referred as adjacency matrix.
Then a normalized spectral clustering algorithm is presented to recover the block structure of adjacency matrix, from which each observation is related to a particular subsystem.
With the labelled observations, parameters of the submodel are estimated via the total least squares (TLS) estimator.
The proposed technique is applicable to switched affine systems with arbitrarily shaped domain partitions, and it offers significantly improved performance and lowered computation complexity than existing techniques.
\end{abstract}


\end{frontmatter}

\section{INTRODUCTION}
\begin{figure}[htb]
  \centering
  \includegraphics[width=\hsize]{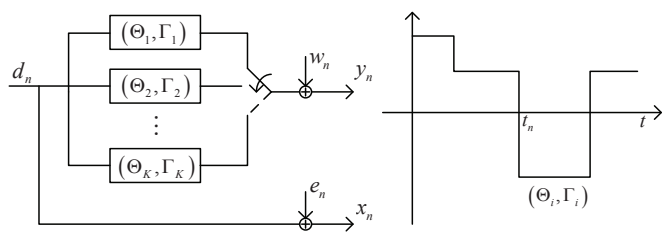}
  \caption{\label{fig1}The structure of switched affine models and its trajectory of evolution. The system can also be viewed as a jumped affine system with a certain
  switching function at the output.}
\end{figure}

We consider the parameter estimation of a switched affine models (SAMs). In its generic formulation, the system switches among a set of submodels. Each of
them can be comprehended as an affine map and is parameterized by two tuples $(\Theta_i,\Gamma_i)$: a linear transformation $\Theta_i$ and a translation $\Gamma_i$.  In particular, if the
system input at time $n$ is switched to $i$th submodel, the system output is determined by parameter matrix $(\Theta_i,\Gamma_i)$. According to the mechanism of
switching, the switched affine systems can be modeled by either piecewise affine models or jump affine models: the former's switch is triggered by its states and
inputs; the latter one switches according to a certain probability.  The switched affine models in this study, regardless of switching mechanism, can be comprehended
as either of them, refer to Fig.~\ref{fig1} for its structure.

Given a set of noisy measurements $\{(x_n, y_n)\}$, we are interested in estimating $(\Theta_i,\Gamma_i)$. If the observations of system are perfectly labelled, then
the problem is one of classical system identification: the subsystem parameter $(\Theta_i,\Gamma_i)$ can be estimated using the corresponding input-output data.  The
challenge, however, is that we do not know the labels of observations a priori.  Thus, the problem of parameter estimation of SAMs is a mixture of two sub-problems: associating observations to subsystems and estimating parameters of submodels.

\subsection{Related Work}  The switched affine models considered in this paper is a simple abstraction of a hybrid system \citep{Heemels}, and its practical significance
in control systems and dynamics \citep{Grieder_2005,Habets_2006}, fault detection and diagnosis \citep{Meskin10TAC,Wu12Auomatica}, pattern recognization\citep{TuragaEtal2008,Vidal2011}, system biology \citep{Guha&Biswas2008},  and wireless communications and networking \citep{Doucet2000AC,Biao&Tong2001} has be well documented.

There is a substantial literature on the identification of switched affine systems \citep{Juloski,Paoletti2007}, including the piecewise affine systems and jump markov linear systems.  According to the knowledge of noise, existing approaches can be summarized into three categories: Gaussian noise, bounded noise and noise with probability density function (\textit{pdf}) given.

Under Gaussian noise assumption (zero mean and unknown variance), the observation association problem can be treated as an unsupervised classification. Since there is no obvious cluster exhibited in input-output space, the observations are usually one-one mapped into a feature parameter space, with an assumption that local data tend to belong the same subsystem \citep{Ferrari}. Then data association is performed via clustering the parameter vectors in feature parameter space; the typical clustering techniques can be adopted in this process, such as $K$-means, $K$-medoids. Fuzzy clustering combined with competitive learning has also been proposed \citep{Gegundez}.

Different with the clustering based methods, algebraic geometric based approaches, also referred as Generalized Principal Component Analysis (GPCA), avoid the data clustering step by imposing a set of constraints to observations \citep{Vidal2005,Ozay2009}. In \citep{Vidal2005}, the authors represent the union of $K$ subspaces in the form of a homogeneous polynomial whose coefficients are estimated in the sense of least squares, referred as \textit{hybrid decoupling constraints} in \citep{Vidal2003}; then system's parameters are calculated by a polynomial derivative algorithm. While the approach is amendable for switched MIMO systems, it is appropriate only for a small number of submodels and parameters. Because the number of coefficients in hybrid decoupling polynomial exponentially increases respect to the numbers of submodels and parameters, this kind of methods are just applicable to SAMs with limited number of submodels.

When the noise is bound, the parameter estimation of SAMs can be formulated as a constrained optimization problem.  The parameters of SAMs are usually optimized in the sense of minimum number of submodels or minimum number of switches \citep{Bemporad2005,Ozay2012}. Thus, the number of subsystems is also estimated in this scenario. Greedy algorithm and some convex optimization techniques, such as semi-definite program \citep{Feng_C} and reweighted $\ell_1$ minimization \citep{Bako2011}, are adopted to solve the best-fitting parameters. Besides, clustering based and algebraic geometric based approaches can also be transplanted to bound noise case. In \citep{Liang2013},  a multi-threshold technique is proposed to associate observations to particular subsystem. A recursive algebraic approach is presented in \citep{Vidal2008} to determine the number and parameters of submodels together.  As the sub-model number determination and parameter estimation are iteratively estimated, these methods tend to perform high computation complexity.

When the probability density function of noise is known, the statistical inference based on classical point estimation and Bayesian estimation methodologies has also been considered.  In particular,  particle filter based Bayesian estimators have been proposed  to obtain the maximum a posteriori (MAP) estimate \citep{Juloski2} or the minimum mean squared error (MMSE) estimate \citep{Doucet2001}. The expectation-maximization  (EM) algorithm has also been applied to estimate parameters of a piecewise affine system  \citep{HayatoNakada} and a
jump markov linear system \citep{Logothetis1999}.  In the field of signal processing, a generalized expectation maximization (GEM) algorithm is used to separate the data sources and determine the parameters of channels \citep{Routtenberg2009}. On the other hand, EM and GEM can be comprehended as data clustering techniques, and classifying them into data clustering based techniques is also reasonable.

The proposed technique belongs to the class of eigen decomposition methods that were popular in the 1990's and remain a significant thrust of estimation methods for system identification.   See, \eg \citep{Viberg:95Automatica,McKelveyAkcayLjung:96IT,TongPerreau:98,An2005} and references therein.
These techniques tend to have ``closed-form'' expressions (in terms of eigen or singular value decompositions).  They typically give perfect identification in the
absence of noise when a certain rank condition is satisfied. When the data are noisy, often there is a signal to noise ratio (SNR) threshold below which the techniques
tend to break down.  The proposed method carries both the strength and weaknesses of these legacy methods.

\subsection{Contributions and organization}

This study is an extension of our previous work, in which parameter estimation of switched linear models (SLMs) is investigated, refer to \cite{Li2013TSP}.  For this paper, the first contribution is estimating the intersection subspace of SAMs from the observations, which can be used to transform SAMs into SLMs. Another novelty is the explicit exploitation of a subspace structure that arises from the switchings among subsystems from one input-output subspace to another. We show that the subspace identified by a singular value decomposition (SVD) provides a natural labeling of input-output observations with the sub-model that generates the observations. The third contribution is the normalized spectral clustering technique proposed to associate observations to subsystems. Once the data labels have been obtained, classical system identification techniques, \eg least squares and total least squares, can then be applied.

Comparing with clustering based methods, the proposed algorithm is applicable to input domains with arbitrarily  shaped (non-convex and possibly non-connected) partitions.
In fact, the proposed technique does not make any limitations for the partitions of input domains.
For standard clustering techniques, it naturally partitions the input domain into convex subregions.  Such a restriction may be unreasonable in practice and is removed here. Another important advantage over non-parametric clustering methods is that,
under the conditions specified in Theorem~\ref{thm1}, the proposed method guarantees perfect identification in the absence of noise.  For clustering based techniques, in contrast, classification errors can happen even if observations are noiseless \citep{Ferrari}. See Section~\ref{sec:simulation} for an example.

When compared with algebraic geometry techniques and statistical inference methods,  the proposed subspace method is based on SVD and numerically more
stable; it does not have issues of convergence or  local optimality. By the way, algebraic geometry techniques recover the systems parameters from the \textit{hybrid
decoupled constraints}, which looks amplifying the effect of noise and introduces a bias to the estimation. This will be shown in the simulation part later.

There is a price paid for the advantages of the proposed approach. A rank condition imposed in Theorem.~\ref{thm1} requires, at the minimum, that the number of observations is higher than the number of submodels.   Not required by existing techniques, this assumption is essential in identifying subspaces associated with each sub-model. For some applications, this assumption may not be acceptable. As already pointed out, the subspace method used here also has the SNR threshold effects. The breakdown threshold depends on the parameters and specific applications. We do not have a characterization of the breakdown threshold. On the other hand, the proposed approach applies for SAMs with non-empty intersection subspace.  This requirement is satisfied in most cases where the number of subsystems is not more than dimension of observations (input and output), and in special cases where the number of subsystems is larger than the dimension of observations, refer to Remark.~\ref{mark1}.

The remainder of the paper is organized as follows: Section II presents the system model and the problem formulation, as well as the assumptions made in
this paper.
Section III presents the key results on a block diagonal structure of the observation subspace.  We present first the idea using the scalar example and then give two
theorems: Theorem~\ref{thm1} describing the full characterization of the observation subspace, and Theorem~\ref{thm3} introducing a way to estimate the intersection subspace of SAMs.
Section IV in further exploits the properties of adjacency matrix, and  proposes a spectral clustering technique that partitions observations into groups, each is associated with a particular subsystem model.  Section V presents  numerical  studies and some comparisons with existed methods. The last section VI gives the conclusions.

Throughout the paper, the following notations are adopted:

\begin{tabular}{|ll|}
\hline
    $N$                                    &the number of observations\\
    $K$                                    &the number of submodels\\
    $d_n$                                  &input of system\\
    $d^{(i)}, D_i$                         &input of $i$th submodel\\
    $y_n, Y_n$                          &observation of output\\
    $x_n, X_n$                          &observation of input\\
    $z_n, Z_n$                          &extended vector of observations\\
    $e_n$                          &input noise\\
    $w_n$                          &output noise\\
    $\theta_i, \Theta_i$                &parameters of $i$th submodel\\
    $\gamma_i, \Gamma_i$                &parameters of $i$th submodel\\
    $S_i$                               &the $i$th observation submodel\\
    $\Omega_i$                          &nominal subspace of $i$th subsystem\\
    $\Omega_0$                          &intersection subspace of subsystems\\
    $\ell_i$                            &label set of $i$th submodel\\
    $A(\Theta)$                         &extended parameters matrix\\
    $V$                                 &subspace of right-singular matrix of $Z$\\
    $M$                                 &adjacency matrix with $M=|VV^{\T}|$\\
    $P$                                 &permutation matrix\\
 \hline
\end{tabular}
\\[0.3em]

\section{Problem formulation}

We consider the discrete-time switched affine models with both input and output noises, refer to Fig.~\ref{fig1} for its structure.  For the $i$th subsystem, its observation can be formulated by the following measurement model
\begin{equation}
  S_i:\left\{  \begin{array}{l}
      x_n^{(i)} = d_n^{(i)}+e_n \\
      y_n^{(i)}=\Theta_i d_n^{(i)} +\Gamma_i+ w_n
    \end{array}\right. \label{eq:model}
\end{equation}
where $d_n^{(i)}$ is the system input and  $x_n^{(i)}\in R^{N_x}, y_n^{(i)}\in R^{N_y}$ are noisy input and output observations of the $i$th subsystem. The observation noise $e_n, w_n$ is
i.i.d. Gaussian with  zero mean and covariance matrix $\mbox{diag}(\sigma_x^2 I, \sigma_y^2 I)$. All input and output variables are vectors, and parameters $(\Theta_i,
\Gamma_i)$ are the linear maps and translation vectors, respectively, with compatible dimensions.

While the model given in (\ref{eq:model})  appears to be a static one, by allowing the input vector $d_n$ to include past observations, the model in fact includes MIMO
systems with finite impulse responses.    When dynamics are included in the modeling, matrices $\Theta_i$ has a block Toeplitz structure that can be exploited.

We also define the nominal or noiseless input-output subspace for the $i$th subsystem
\begin{equation}
    \Omega_i\defeq\left\{(x,y)\left|y=\Theta_i x +\Gamma_i, x \in R^{N_x}, y \in R^{N_y} \right. \right\}, \label{eq:Omega}
\end{equation}
and the intersection subspace for all subsystems
\begin{equation}
  \Omega_0\defeq\bigcap_{i=1}^{K}\Omega_i.
\end{equation}

\begin{figure}[htb]
  \centering
  \includegraphics[width=0.8\hsize]{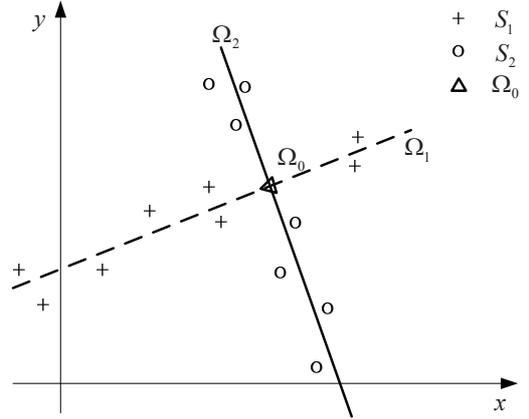}
  \caption{\label{fig:Omega0}The nominal subspace and its observations in 2 dimension case.}
\end{figure}

In noiseless case, the intersection subspace contains elements that satisfy all subsystems' measurement models in (\ref{eq:model}). Observations from intersection subspace should be treated as inherently undecided data. But we will show they are very helpful for data association of switched affine systems.  For a SAMs defined in $R^2$, its intersection subspace is the crossing point of two submodel lines, refer to Fig.~\ref{fig:Omega0} for an illustration.

Given a set of unlabelled input-output measurements $(x_n, y_n)$, which is a mixture of noisy observations of $\{\Omega_i\}_{i=1}^K$, we are interested in associating them to corresponding subsystems and estimating $(\Theta_i,\Gamma_i), i=1,\ldots, K$.

\begin{remark}
\label{mark1}
In this study, we assume that the intersection subspace of SAMs is non-empty. It should be noted that this assumption is hardly suitable when the number of subsystems is larger than the dimension of observations, i.e. $K>N_x+N_y$, or when the SAMs with the same linear transformations, i.e. $\Theta_1=,\ldots,=\Theta_K$.
\end{remark}
\begin{remark}
\label{mark0}
We assume that the number $K$ of submodels  and the dimensions of submodels are fixed and known.   This is an assumption that most existing techniques make and is quite
restrictive. Identifying the number of submodels and their respective dimensions are challenging although there are many practical techniques including some based on
eigenvalue decompositions.
\end{remark}

\section{Subspace Structure of Switching Dynamics}
We present in this section a subspace structure in the observation that shows the decomposition for data associated with different sub-models.

\subsection{A sketch of ideas}
The subspace structure can be easily described using the following simple noiseless {\em scaler} switched affine models:
\begin{align}
    S_1: &\left\{\begin{array}{l}
                      x_n^{(1)} = d_n^{(1)}\\
                      y_n^{(1)}=\theta_1 d_n^{(1)} +\gamma_1
                    \end{array}
             \right.   \label{eq:simple_a}\\
    S_2: &\left\{\begin{array}{l}
              x_n^{(2)} = d_n^{(2)} \\
              y_n^{(2)}=\theta_2 d_n^{(2)} +\gamma_2
            \end{array}
     \right.   \label{eq:simple_b}
\end{align}
where submodels $S_1$ and $S_2$ can be treated as arbitrary non-paralleled lines on real plane.

Let $(x_0,y_0)\in \Omega_0$ be the elements of intersection space  of two submodels. Then it must satisfy both submodels simultaneously, i.e.
\begin{equation}
y_0=\theta_1 x_0+\gamma_1=\theta_2 x_0+\gamma_2
\end{equation}

Then we can rewrite the bi-model SAMs into a switched linear models as follows
\begin{align}
    S(\Omega_1-\Omega_0): &\left\{\begin{array}{l}
                      x_n^{(1)}-x_0=d_n^{(1)}-d_0\\
                      y_n^{(1)}-y_0=\theta_1 (d_n^{(1)}-d_0)
                    \end{array}
             \right.   \label{eq:simple_a1}\\
    S(\Omega_2-\Omega_0): &\left\{\begin{array}{l}
              x_n^{(2)}-x_0=d_n^{(2)}-d_0 \\
              y_n^{(2)}-y_0=\theta_2 (d_n^{(2)}-d_0)
            \end{array}
     \right.   \label{eq:simple_b1}
\end{align}

Consider, for the moment, the special case when the system input $d_n$ arrives sequentially, first $N_1$ samples from sub-model $1$ and the next $N_2=N-N_1$ from
sub-model $2$. We collect input-output data in a matrix
\begin{align}
  Z &\defeq \left[\begin{array}{ccc}
    x_1 & \cdots & x_N\\
    y_1 & \cdots & y_N
  \end{array}\right],\\
  Z_0 &\defeq \left[\begin{array}{ccc}
    x_0 & \cdots & x_0\\
    y_0 & \cdots & y_0
  \end{array}\right]\label{eq:Z0}
\end{align}
The switched affine system then has a linear form
\begin{equation} \label{eq:Z=AU}
Z-Z_0=A\left(D^{\T}-D_0^{\T}\right),
\end{equation}
where
\begin{eqnarray}
A &\defeq & \left[\begin{array}{cc}
  1 & 1\\
    \theta_1&\theta_2\\
  \end{array}\right], \\
D_0^{\T} & \defeq &  \left[\begin{array}{cccccc}
    d_0&\ldots&d_0&0&\ldots&0\\
    0&\ldots&0&d_0&\ldots& d_0
  \end{array}\right]\label{eq:U0}\\
D^{\T} & \defeq &  \left[\begin{array}{cccccc}
    d_1^{(1)}&\ldots&d_{N_1}^{(1)}&0&\ldots&0\\
    0&\ldots&0&d_{N_1+1}^{(2)}&\ldots& d_{N}^{(2)}
  \end{array}\right]\nonumber\\
&=&  \mbox{diag}(D_1^{\T}, D_2^{\T})  \label{eq:diagU}
\end{eqnarray}
where vector $D_1^{\T}=[d_1^{(1)},\cdots, d_{N_i}^{(1)}]$ from the first sub-model
 with parameter $(\theta_1,\gamma_1)$ and the next $N_2$ samples form vector  $D_2^{\T}$  from the second sub-model with parameter $(\theta_2,\gamma_2)$.

In practice, the system input sequence is of course not ordered as such, and matrix $D^{\T}$ is post-multiplied by some (unknown) permutation matrix $P$.  We will deal
with the presence of unknown permutation $P$ in the subsequent formal characterization of the subspace decomposition.

Let $Z-Z_0$ have the singular value decomposition of the form
\begin{equation}
  \label{SVD}
  Z-Z_0=Q\Sigma V^{\T},
  \end{equation}
Note that $\Sigma>0$ is a diagonal matrix with reduced dimension $2\times 2$, $Q$ is an orthogonal matrix, and
 $V^{\T}$ is a $2\times N$ matrix of the same dimension of input matrix $D^{\T}$.  Indeed, when $A$ is nonsingular,
$V^{\T}$ spans the same row space as $D^{\T}$.  In other words, we have
\begin{equation} \label{eq:VT=TU}
V^{\T}=T\left(D-D_0\right)^{\T},~~T\defeq Q^{\T}\Sigma^{\text{\tiny -1}}A.
\end{equation}
Because $V^{\T}V=T\left(D-D_0\right)^{\T}\left(D-D_0\right)T^{\T}=I$ and $T$ is nonsingular,
\begin{equation} \label{eq:TT}
T^{\T}T=\left[\left(D-D_0\right)^{\T}\left(D-D_0\right)\right]^{\text{\tiny -1}}.
\end{equation}
Thus, we have
\begin{align}\label{eq:VVT}
VV^{\T} &= \left(D-D_0\right)T^{\T}T\left(D-D_0\right)^{\T}\notag\\
        &=\left(D-D_0\right)\left[\left(D-D_0\right)^{\T}\left(D-D_0\right)\right]^{\text{\tiny -1}}\left(D-D_0\right)^{\T}\notag\\
        &= \left[\begin{array}{cc}
                  \frac{(D_1-d_0)(D_1-d_0)^{\T}}{||D_1-d_0||^2} & 0 \\
                  0 & \frac{(D_2-d_0)(D_2-d_0)^{\T}}{||D_2-d_0||^2} \\
                \end{array}
                \right].
\end{align}
Here, misuse the symbols for the moment, $D_1-d_0$ represents subtracting $d_0^{\T}$ from each row of $D_1$, so does $D_2-d_0$. Thus, matrix  $VV^{\T}$ is block
diagonal with the first block made of data from only the first sub-model  and second block from only the second sub-model. The key observation is that matrix $VV^{\T}$
gives the partition of input sequence which is exploited in the proposed algorithm.

If the system input sequence is not arranged in the block sequential fashion as in (\ref{eq:diagU}), matrix $VV^{\T}$ is no longer block diagonal; instead it will be a
block diagonal matrix under a certain similar transform defined by a permutation matrix $P$.  However, the permutation induced transform can be easily reversed to
recover the block diagonal structure.

\subsection{Subspace decomposition}
We now formally state the subspace decomposition by removing the sequential arrival restriction on the input sequence.
For a $K$-submodel switched affine systems, define
\begin{align}
  \mbox{Observations:}\quad &z_n\triangleq\left[\begin{array}{c}
                                      x_n \\
                                      y_n
                                    \end{array}
  \right], Z\triangleq\left[z_1, \ldots, z_N\right]\\
  \mbox{Parameter matrix:}\quad &A(\Theta)\triangleq\left[\begin{array}{ccc}
                                                 I & \ldots & I \\
                                                 \Theta_1 & \ldots & \Theta_K
                                               \end{array}
  \right]\\
  \mbox{Input matrix:}\quad &D^{\T}\triangleq \mbox{diag}\left(D_1^{\T}, \ldots, D_K^{\T}\right)
\end{align}
where $D_i^{\T}$ is the matrix of ordered (according to the time of arrival) input vectors of the $i$th subsystem.

In the absence of noise, the system can be formulated in a linear model as
\begin{equation}
  \label{Z=AUP}
  Z-Z_0=A(\Theta)\left(D-D_0\right)^{\T}P
\end{equation}
where $Z_0$ is a rank-one matrix constructed by elements of intersection space in form of (\ref{eq:Z0}), $D_0$ is the input matrix of intersection subspace constructed
in form of (\ref{eq:U0}), and $P$ is an unknown permutation matrix that permute the ordered input sequence to the actual sequence of arrivals.

The following theorem captures the subspace structure in the row space of data matrix $Z-Z_0$.
\begin{theorem}\label{thm1}
Consider the noiseless system model given in (\ref{Z=AUP}). Assume that $A(\Theta)$ has full column rank and the intersection space of subsystems $\Omega_0$ is
non-empty.  Let $Z-Z_0$ have the singular value decomposition of the form
  \begin{equation}
    \label{SVD}
    Z-Z_0=Q\Sigma V^{\T}
  \end{equation}
  where $QQ^T=I$ and $V^{\T}$ is of the same size as $D^{\T}$ with orthogonal rows. Then
  \begin{equation}
  \label{Lambda}
    VV^{\T}=P\mbox{diag}\left(\Lambda_1,\ldots, \Lambda_K\right)P^{\T},
  \end{equation}
  where $\Lambda_i=\left(D_i-D_0\right)\left[\left(D_i-D_0\right)^{\T}\left(D_i-D_0\right)\right]^{\text{\tiny -1}}\left(D_i-D_0\right)^{\T}$.
\end{theorem}

\vspace{0.5em}
\begin{pf}
See Appendix.~A.
\end{pf}

\vspace{0.5em}
\begin{remark}
If the system input sequence arriving in consecutive from the same sub-model, we have $P=I$.  In this case, matrix $VV^{\T}$ is block diagonal. In general, matrix $P$
scrambles $\mbox{diag}(\Lambda_1,\cdots, \Lambda_K)$.  For noiseless measurement, the matrix can be easily de-scrambled by permuting $VV^{\T}$ back to diagonal form.
See \citep{Aykanat&Pinar&Catalyurek:04SIAM} for an efficient de-permutation algorithm.  But the problem at hand is more challenging due to the presence of noise.
\end{remark}

\begin{remark}  In the presence of noise, matrix $VV^{\T}$ no longer has the (permutation transformed) block diagonal form in general.  Simulation at various noise
levels suggest that the structure existed in the noiseless case holds approximately.
\end{remark}

\begin{remark} The assumption that $A(\Theta)$ has full column rank is restrictive.  As the number of sub-models increases, the number of required independent output
observations must also increase.  This puts a physical limitation on the type of system to which the algorithm considered here is applicable.
\end{remark}

\begin{remark} We assume that the input matrix $D^{\T}$ has full row rank and each sub-block $(D_i-D_0)^{\T}$ cannot be row-permutated into a diagonal form.
The full rank condition on $D$ requires that the each subsystem is persistently excited---a necessary condition for identifiability assuming that we can associate
observation data samples $\{z_n\}$ perfectly with the submodels $\{\Theta_i\}$ that generates them.

The condition on individual input $(D_i-D_0)^{\T}$ not permutable to diagonal form is also necessary for identifiability.   If $(D_i-D_0)^{\T}=\mbox{diag}(D_{11}^{\T},
D_{12}^{\T})$, then we can break up $A(\Theta_1)$ accordingly into $A(\Theta_{11})$ and $A(\Theta_{12})$ and group $(\Theta_{12},D_{12}^{\T})$ with the second submodel
$(\Theta_2,D_2^{\T})$.  The system then becomes unidentifiable.
\end{remark}

\subsection{Estimation of intersection subspace}

In the content above, the intersection subspace is assumed as known. In this part, we propose a theorem to determine it from the observations of system. Let's introduce
a  homogeneous polynomial constructed by the parameters of switched affine models
\begin{equation}
\label{eq:H}
  H(x,y)\triangleq\prod_{i=1}^K(y-\Theta_ix-\Gamma_i).
\end{equation}
Where column vector $x, y$ has the same dimension with $x_n, y_n$, $\Theta_i, \Gamma_i$ corresponds to the parameters of $i$th submodel.

In noiseless case, it is easy to verify following equation identically holds for all observations,
\begin{equation}
  H(x_n,y_n)=\prod_{i=1}^K(y_n-\Theta_ix_n-\Gamma_i)=0,n=1,\ldots,N \label{eq:hdc}
\end{equation}
This equation is referred as \textit{Hybrid Decoupled Constraint} in \citep{Vidal2003}.

In noiseless case, the coefficients of $H$ can be solved directly when the number of observation is larger than its order. Moreover, the the parameters of submodels can
be recovered from the coefficients of $H$ in this scenario, refer to Vidal's work in \citep{Vidal2003}.  When the observation data are corrupted by noise, the
coefficients of $H$ need to be determined in the sense of least squares technique. Although the parameters can be recovered by the GPCA technique in \citep{Vidal2005}, we will illustrate in the simulation that it introduces bias to the estimations of the parameters.

We know that elements of intersection subspace satisfies all submodels' models. So they must be the $K$-fold multiple roots of polynomial $H$ in (\ref{eq:hdc}). We
propose following theorem to determine the non-empty intersection subspace of switched affine systems.
\begin{theorem}\label{thm3}
Consider the switched affine models defined in (\ref{eq:model}). In noiseless case, its not-empty intersection subspace can be solved by following equations
  \begin{equation}\label{eq:Omega0}
    \Omega_0\defeq\left\{(x_0,y_0)\left|
                                          \left.\frac{\partial^{\text{\tiny K-1}}H}{\partial^{\text{\tiny K-1}}x}\right|_{x=x_0,y=y_0}=0,
                                          \left.\frac{\partial^{\text{\tiny K-1}}H}{\partial^{\text{\tiny K-1}}y}\right|_{x=x_0,y=y_0}=0
    \right.\right\}.
  \end{equation}
  where $H$ is the homogeneous polynomial defined in (\ref{eq:H}), $\partial^{\text{\tiny K-1}}H/\partial^{\text{\tiny K-1}}$ represents the $(K-1)$ order partial
  derivative of $H$.
\end{theorem}

\vspace{0.5em}
\begin{pf}
The proof can be easily derived from the properties of multiple roots of polynomial, hence omitted here.
\end{pf}

\begin{remark}
For the MIMO case, equations in (\ref{eq:Omega0}) maybe over determined, the intersection subspace can be fixed by solving compatible number of equations, or
approximated by solving all equations in the sense of least squares.
\end{remark}
\begin{remark}
For the switched affine systems with empty intersection space, equations in (\ref{eq:Omega0}) are still solvable, but the solutions in this scenario are useless.
\end{remark}

\section{Spectral clustering on subspace and parameter estimation\label{sec:spectral}}
With the theorems above, the data association problem turns into one of recovering the block diagonal structure $\mbox{diag}(\Lambda_1,\ldots, \Lambda_K)$
from $VV^{\T}$. It also can be understood as identification of permutation matrix $P$. Once the labels of observations are obtained, the problem of estimating system
parameters become standard. In this section, we present  a spectral clustering on subspace (SCS) algorithm that recovers the block diagonal structure in the observation
subspace from $VV^{\T}$.  Spectral clustering technique was originally proposed for the purpose of graph-cutting, refer to \citep{Luxberg07SC,ShiMalik2000} and
references therein. However, the problem at hand has some useful properties that enable us to develop more efficient algorithm.

\subsection{The properties of $VV^{\T}$}
Let's start with the scaler case defined in (\ref{eq:simple_a}) and (\ref{eq:simple_b}) and introduce an \textit{adjacency matrix} defined by
\begin{equation}
\label{eq:M}
  M=|VV^{\T}|.
\end{equation}
Here $\mid\cdot\mid$ represents the element-wise operation of taking absolute values of entries of the matrix.  Because a permutation matrix $P$ acts on $VV^{\T}$
symmetrically, to de-permutate the matrix $M$ is equivalent to recovering the order of its diagonal elements. From the expression of $VV^{\T}$ and $U$ in
(\ref{eq:VVT}), we observe following useful corollary.
\begin{corollary}
\label{cor}
  In noiseless case, for the adjacency matrix defined in (\ref{eq:M}) and inputs excluding elements of intersection space,
  \begin{enumerate}[(1)]
    \item \label{cor1}if the element $M(m,n)\neq 0$, then systems' inputs $d_m$ and $d_n$ belong to the same subsystems. Specifically, the element of adjacency matrix
        $M(m,n)$ can be expressed as
         \begin{equation}\label{M_ij}
         M(m,n)=\frac{(d_m-d_0)(d_n-d_0)^{\T}}{||D_i-d_0||^2}.
         \end{equation}
         Where $d_0$ is from the intersection space, $D_i$ is the input matrix of $i$th subsystem.
    \item \label{cor2}if the element $M(m,n)= 0$, then system's inputs $d_m$ and $d_n$ are from different subsystems.
  \end{enumerate}
The inverse propositions of (\ref{cor1}) and (\ref{cor2}) still hold. Note that when system's inputs include $d_0$, the proposition (\ref{cor2}) does not hold any more.
\end{corollary}

In the noiseless case, the block diagonal structure of $M$ can be de-permutated by comparing its off-diagonal elements. However, the corollary~\ref{cor} above barely
holds in vector case, we need to develop a more robust technique.

\subsection{Spectral clustering on subspace\label{sec:alg}}

For the adjacency matrix $M$, following theorem further exploits the block diagonal structure in its singular value decomposition.
\begin{theorem}\label{thm2}
Consider the noiseless system model given in (\ref{Z=AUP}) and adjacency matrix $M=|VV^{\T}|$ in (\ref{eq:M}). Let $M$ be normalized as
  \begin{align}
    \bar{M}&=W^{\tiny -1/2}MW^{\tiny -1/2}\\
    W&\triangleq diag(\sum_{j=1}^Nm_{1,j}, \ldots, \sum_{j=1}^Nm_{N,j})
  \end{align}
Then $\bar{M}$ has $K$ largest singular values
\begin{equation}
  \delta_1=, \ldots, \delta_K=1,
\end{equation}
and their corresponding eigenvectors come from the $K$ subsystems respectively. The singular value decomposition has the form of
\begin{align}
  \bar{M}&=EGE^{\T}\\
  G&=diag(\overbrace{1, 1, \ldots, 1}^K, *)\\
  E&=Pdiag(E_1, \ldots, E_K, *)
\end{align}
 Where $P$ is the unknown permutation matrix and $E_i$ is the eigenvector of normalized $\Lambda_i$ defined in theorem.~\ref{thm1}.
\end{theorem}

\vspace{0.5em}
\begin{pf}
Let's denote the block diagonal structure in (\ref{Lambda}) as follows
\begin{equation}
  \label{permu_W}
   \Lambda=diag(\Lambda_{1}, \Lambda_{2}, \ldots, \Lambda_{K})
\end{equation}
where $\Lambda_{i}=\left(D_i-D_0\right)\left[\left(D_i-D_0\right)^{\T}\left(D_i-D_0\right)\right]^{\text{\tiny -1}}\left(D_i-D_0\right)^{\T}$.

Normalize $\Lambda_i$ as
  \begin{align}
    \bar{\Lambda}_i&=\Sigma_i^{\tiny -1/2}\Lambda_i\Sigma_i^{\tiny -1/2}\label{bar_Wi}\\
    \Sigma_i&\triangleq diag(\sum_{j=1}^{N_i}\lambda_{1,j}, \ldots, \sum_{j=1}^{N_i}\lambda_{N_i,j})
  \end{align}
Here $\lambda_{*,j}$ is the element of $\Lambda_i$, $\Sigma_i^{\tiny -1/2}$ is the element-wise operation. Then $\bar{\Lambda}_i$ can be expressed in form of
\begin{align}
  &\bar{\Lambda}_i\mu_i^{\tiny 1/2}=1\times \mu_i^{\tiny 1/2}\\
  &\mu_i=\left[\sum_{j=1}^{N_i}\lambda_{1,j}, \ldots, \sum_{j=1}^{N_i}\lambda_{N_i,j}\right]^{\T}
\end{align}
This means that $\bar{\Lambda}_i$ has the eigenvalue $\delta=1$ and its corresponding eigenvector is $\mu_i^{\tiny 1/2}$ (note that $\mu_i^{\tiny 1/2}$ is not normalized here).

According to the definition of $\bar{\Lambda}_i$ in (\ref{bar_Wi}), we have
\begin{equation}
  \rho(\bar{\Lambda}_i)\leq\|\bar{\Lambda}_i\|_1 \leq 1.
\end{equation}
Here $\rho(\bar{\Lambda}_i)$ is the spectral radius of $\bar{\Lambda}_i$ and $\|\bar{\Lambda}_i\|_1$ is the induced 1-norm of $\bar{\Lambda}_i$. Thus, $\delta=1$ is the largest
eigenvalue of $\bar{\Lambda}_i$.

For the block diagonal matrix $\Lambda$, define
  \begin{align}
    \bar{\Lambda}&=diag(\bar{\Lambda}_1, \ldots, \bar{\Lambda}_K)=\Sigma^{\tiny -1/2}\Lambda\Sigma^{\tiny -1/2}\\
    \Sigma&\triangleq diag(\Sigma_1, \ldots, \Sigma_K)
  \end{align}

Naturally, the largest $K$ eigenvalues of $\bar{\Lambda}$  all equal to $1$ and the corresponding eigenvectors of $\bar{\Lambda}$ are the eigenvectors of
$\bar{\Lambda}_i$ with the positions of the other blocks filled with $0$. That is
\begin{equation}
  \tilde{\mu}_{\tiny [1:K]}=\begin{bmatrix}
   \mu_{1}^{\tiny 1/2}&0&\cdots&0\\
   0&\mu_{2}^{\tiny 1/2}&\cdots&0\\
   0&0&\ddots&0\\
   0&0&\cdots&\mu_{\tiny K}^{\tiny 1/2}
 \end{bmatrix}.
\end{equation}

For the permutated matrix $M=P\Lambda P^{\T}$, define
\begin{equation}
  W\triangleq diag(\sum_{j=1}^Nm_{1,j}, \ldots, \sum_{j=1}^Nm_{N,j})=P\Sigma P^{\T}
\end{equation}
Then
\begin{equation}
  \bar{M}=W^{\tiny -1/2}MW^{\tiny -1/2}=P\Sigma^{\tiny -1/2}P^{\T}P\Lambda P^{\T}P\Sigma^{\tiny -1/2}P^{\T}=P\bar{\Lambda}P^{\T}
\end{equation}

Thus, $\bar{M}$ has the same largest $K$ singular value with $\bar{\Lambda}$: $\delta_1=, \ldots, =\delta_K=1$ and the singular value decomposition in form of
\begin{equation}
  \bar{M}=EGE^{\T}
\end{equation}
  \begin{equation}
  G=\begin{pmat}[{|}]
    {\begin{array}{cccc}
                 1 & & &  \\
                 & 1 & &  \\
                 & & \ddots &  \\
                 & & & 1
               \end{array}}&0\cr\-
    0&*\cr
  \end{pmat}, E=P\begin{pmat}[{|}]
    {\begin{array}{cccc}
      E_1& 0 & \ldots&0 \\
       0& E_2& \ldots&0 \\
       0&  0 & \ddots&0 \\
       0&  0 & \ldots&E_K
    \end{array}} &*\cr
  \end{pmat}.
 \end{equation}
where $E_i=\mu_{i}^{1/2}$ is the eigenvector of $\bar{\Lambda}_i$.
\end{pf}

\vspace{0.5em}

\begin{figure}[!h]
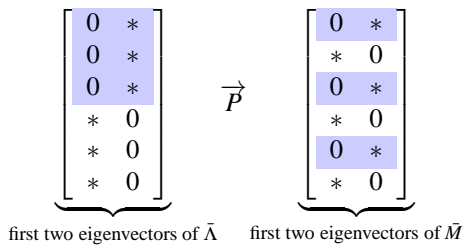

\centering
\begin{displaymath}
\underbrace{\left[\begin{array}{cc}
                \cellcolor{blue!20}0 & \cellcolor{blue!20}* \\
                \cellcolor{blue!20}0 & \cellcolor{blue!20}* \\
                \cellcolor{blue!20}0 & \cellcolor{blue!20}* \\
                * & 0 \\
                * & 0 \\
                * & 0
              \end{array}\right]
  }_{\text{first two eigenvectors of~} \bar{\Lambda}}\overrightarrow{P}\underbrace{\left[\begin{array}{cc}
                \cellcolor{blue!20}0 & \cellcolor{blue!20}* \\
                * & 0 \\
                \cellcolor{blue!20}0 & \cellcolor{blue!20}* \\
                * & 0 \\
                \cellcolor{blue!20}0 & \cellcolor{blue!20}* \\
                * & 0
              \end{array}\right]
  }_{\text{first two eigenvectors of~} \bar{M}}
\end{displaymath}
\caption{\label{W_M}The rows structure of eigenvectors of $\bar{\Lambda}$ and $\bar{M}$. The rows in shadow belong to the sub-model one and the others belong to
sub-model two.}
\end{figure}

In the theorem above, we can find that the permutation structure of measurement is reflected in the first $K$ eigenvectors of $\bar{\Lambda}$ and only the rows of
eigenvector are permutated in the SVD of $\bar{M}$. It's very easy to recover the block diagonal structure of $M$ according to the position of $0$-elements in the first
$K$ eigenvectors of $\bar{M}$. For example, in Fig.~\ref{W_M}, a bi-model SAMs with 6 inputs (3 from subsystem one and another 3 from subsystem two).

In the absence of noise, the $K$ eigenvectors are indicator functions that directly associate each observation to a particular subsystem.  With noise disturbance, these
eigenvectors hardly hold zero entries any more and the structure in Fig.~\ref{W_M} will become indistinct.  In this scenario, each row of $K$ eigenvectors is taken as a
joint indicator for subsystems, then all rows are grouped into $K$ clusters by $K$-means. Intuitively speaking,  $K$ eigenvectors contain redundant information for data
association, which restrains the disturbance of noise. On the other hand, since $K-1$ eigenvectors are necessary to fix the structure of observations, there is little
computation amount increase to cluster $K$ eigenvectors in noise case.

\begin{figure*}[thpb]
  \centering
  \includegraphics[width=0.48\hsize]{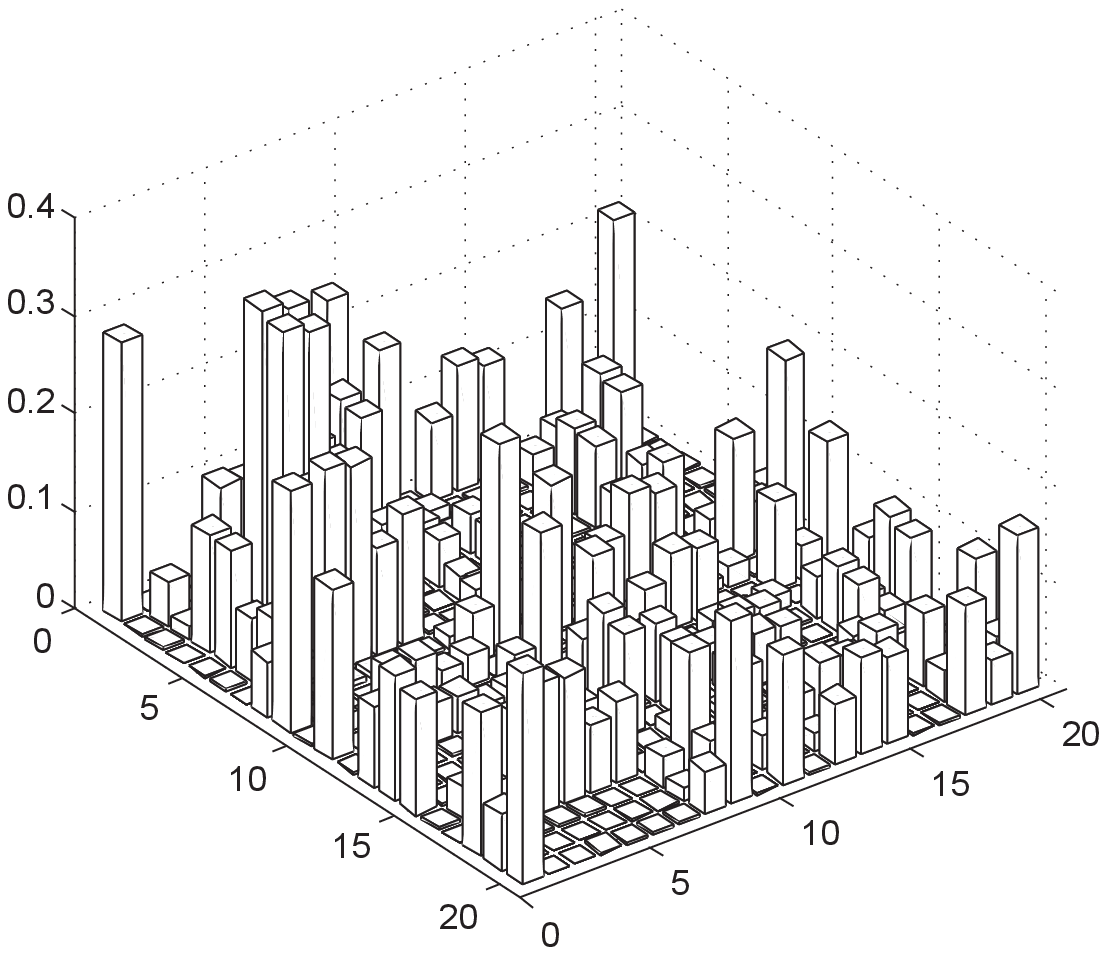}
  \includegraphics[width=0.48\hsize]{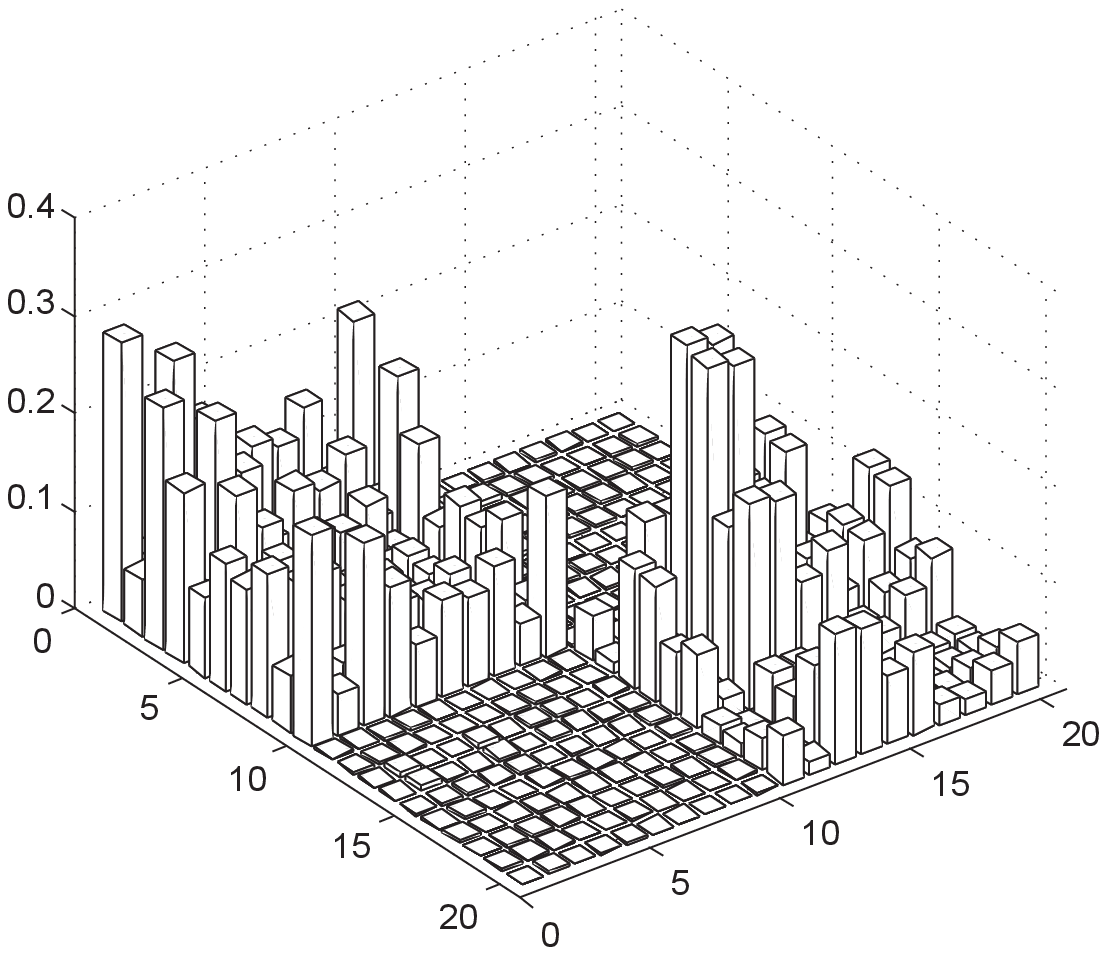}
  \caption{\label{fig_M_W}3-D bar graphs for a $20 \times 20$ matrix $M=|VV^{\T}|$ and its de-permutated result.}
\end{figure*}

We propose a spectral clustering on subspace (SCS) algorithm to recover the block diagonal structure of $M$ in Algorithm.~\ref{alg1}: Step 1 of the algorithm calculate
the intersection subspace according to the Theorem~\ref{thm3}; Step 2 computes the best row space of the observation matrix $Z-Z_0$ in the form of $V^T$, and the
adjacency matrix $M=|VV^{\T}|$; Step 3 normalizes the adjacency matrix $M$ into $\bar{M}$; Step 4 does the SVD on $\bar{M}$, or eigenvalue decomposition as $\bar{M}$ is
symmetrical; In Step 5, the first $K$ eigenvectors of $\bar{M}$ are clustered into $K$ groups by $K$-means, where $\ell_i$ is the label set for the $i$th sub-model;
note that, if do eigenvalue decomposition on $\bar{M}$ in step 4, make sure that the first $K$ eigenvectors of $\bar{M}$ correspond to the largest $K$ eigenvalues.
Step 6 gives the parameters' estimation via total least squares technique.
%For input-output noisy case, also referred as Errors-in-variables models,  there are many well studied identification methods, especially for the dynamic models, refer to \cite{Soderstrom2007} and references therein. In this study, we just adopt total least squares technique to get the parameters' estimation.
%Step 6 adopts a post processing to refine the estimations, and makes the output error achieve a local minimum.
To illustrate the result of de-permutation, Fig.~\ref{fig_M_W} shows a $20\times 20$ matrix de-permutated by the proposed algorithm.

%%%%%%%%%%%%%%%%%%%%%%%%%%%%%%%%%%%%%%%%%%%%%%%%%%%%%%%%%%%%%%%%%%%%%%%%%%%%%%%%%%%%%%%%%%%%%%%%%%%%%%%%%%%%%%%%%%%%%%%%%%%%%%%%%
\begin{algorithm}
  \caption{\label{alg1} Spectral clustering on subspace (SCS) algorithm.}
  \begin{algorithmic}
  \REQUIRE \STATE Observations $X=\{x_n\}_{n=1}^{\N},Y=\{y_n\}_{n=1}^{\N}$, $Z=[X;Y]$
  \ENSURE \STATE Parameter estimations $\{\hat{\Theta}_i,\hat{\Gamma}_i\}_{i=1}^{\K}$
    \STATE (1) calculate the intersection subspace $\Omega_0$, and denote $Z_0=\left[x_0^T,y_0^T\right]^T\in\Omega_0$
    \STATE (2) singular value decomposition on measurements $Z-Z_0$
    \STATE    \qquad\qquad$Z-Z_0=Q\Sigma V^{\T}$
    \STATE    \qquad note that $V^T$ has the same size of $Z$
    \STATE (3) normalize the adjacency matrix $M=|VV^{\T}|$
    \STATE     \qquad\qquad$\bar{M}=W^{\tiny -1/2}MW^{\tiny -1/2}$
    \STATE     \qquad\qquad$W\triangleq diag(\sum_{j=1}^Nm_{1,j}, \ldots, \sum_{j=1}^Nm_{N,j})$
    \STATE (4) singular value decomposition on $\bar{M}$
    \STATE     \qquad\qquad$\bar{M}=EGE^{\T}$
    \STATE (5) cluster the rows of first $K$ eigenvectors with $K$-means
               \STATE\qquad $[cent, ind]=kmeans(E(:,1:K),K)$
               \STATE\qquad\qquad $\ell_1=find(ind==1)$
               \STATE\qquad\qquad $\ell_2=find(ind==2)$
               \STATE\qquad\qquad \ldots
               \STATE\qquad\qquad $\ell_{\N}=find(ind==K)$
    \STATE (6) estimate the parameters with total least squares (\textit{tls})
    \STATE     \qquad\qquad$\left[\hat{\Theta}_i,\hat{\Gamma}_i\right]=tls(X(\ell_i),Y(\ell_i)), i=1,\ldots, K$
%    \STATE     \qquad\qquad $Eout_n=\|y_{i,n}-\hat{\Theta}_ix_{i,n}\|, n=1,\ldots, N$
%    \STATE (6) reclassify and update the estimations
%                \REPEAT
%    \STATE       $[Eout,Ind]=sort(Eout,1,'descend')$
%                    \FOR{$j=1\to N$}
%                    \STATE reclassify the $Ind(j)$th data
%                    \IF {$\sum_{n=1}^{\N} Eout_n$ decreases}
%                        \STATE update label set $\ell_i$, $\hat{\Theta}_i$ and $Eout_n$
%                    \ENDIF
%                    \ENDFOR
%                \UNTIL{$\sum_{n=1}^{\N} Eout_n$ no longer decreases}
  \RETURN $\{\hat{\Theta}_i,\hat{\Gamma}_i\}_{i=1}^{\K}$
\end{algorithmic}
\end{algorithm}
\begin{remark}
   According to the property of $VV^{\T}$ in (\ref{M_ij}), inputs near the intersection subspace produce small elements in $M$ which cause confusion with zero. In the
   estimation step, one option is to cut off the small diagonal elements of $M$ via a threshold, the other one is to adopt a total least squares weighted by the
   diagonal elements of $M$.
\end{remark}

\section{Simulation examples\label{sec:simulation}}
In this section, we present numerical studies and compare the proposed algorithm, referred as SCS algorithm, with two benchmark techniques.  One benchmark technique is  based on the $K$-means
clustering algorithm \citep{Ferrari}, the other is an algebraic geometric technique referred as Generalized Principal Component Analysis (GPCA)  in \citep{Vidal2005}. $K$-means estimator is an iterative scheme. To avoid local optimal solution, the implementation, in this simulation, repeats the clustering process multiple times, each with a new set of initial centroids. We also compare the performance of various estimators using ``clairvoyant'' Maximum Likelihood  (C-ML) where we assume that labels of observations are known.

We use the standard mean squared error (MSE) as the performance metric in comparing different estimators.  Here the statistical average is taken with respect to random
measurement noise in the system input and output.
In Monte Carlo simulations, sample MSE is used as an estimate of the actual MSE. The system input sequence $d_n$ is considered deterministic and is fixed in the Monte
Carlo simulations.

We are interested in the performance of different estimators at different levels of signal-to-noise ratio (SNR). In particular, we define SNR (in dB) as
\begin{equation}
\mbox{SNR}=10 \log  \frac{\sum_{n=1}^N  \left(\sum_{i=1}^K ||\Theta_id_n+\Gamma_i||^2 \times \pi_{n,i}+ ||d_n||^2\right)}{N(N_x\sigma_e^2+N_y\sigma_w^2)},
\end{equation}
where $\pi_{n,i}\in \{0,1\}$ is an indicator for submodel $i$ which takes $1$ only if the $d_n$ is applied to subsystem $i$ and $0$ otherwise. In the denominator, $N_x$
is the dimension of vector $x_n$ and $N_y$ the dimension of $y_n$.  Intuitively, the numerator is the total signal energy in the input and out sequence and the
denominator the total noise energy.

\subsection{SISO piecewise affine models with two submodels}
The first numerical case was
 an bi-model piecewise affine models with scalar parameters. For each submodel,
\begin{align}\label{case1}
    S_1: &\left\{\begin{array}{l}
                      x_n = d_n+e_n\\
                      y_n=1.7d_n+0.9+w_n
                    \end{array}
             \right., d_n\geq 0  \notag\\
    S_2: &\left\{\begin{array}{l}
              x_n = d_n \\
              y_n=2.8d_n+1.2+w_n
            \end{array}
     \right., d_n < 0  \notag
\end{align}
Where $w_n\sim N(0,\sigma_1^2)$, $e_n\sim N(0,\sigma_2^2)$ are i.i.d. random sequences. Similar examples were used in \citep{Ferrari,Vidal2003} except that we have made
the scenario more challenging by making the parameters of the two subsystems relatively close. In the simulation, the input data were generated by standard normal
distribution and kept fixed in Monte Carlo runs.

\begin{figure*}[thpb]
  \centering
  \includegraphics[width=0.48\hsize]{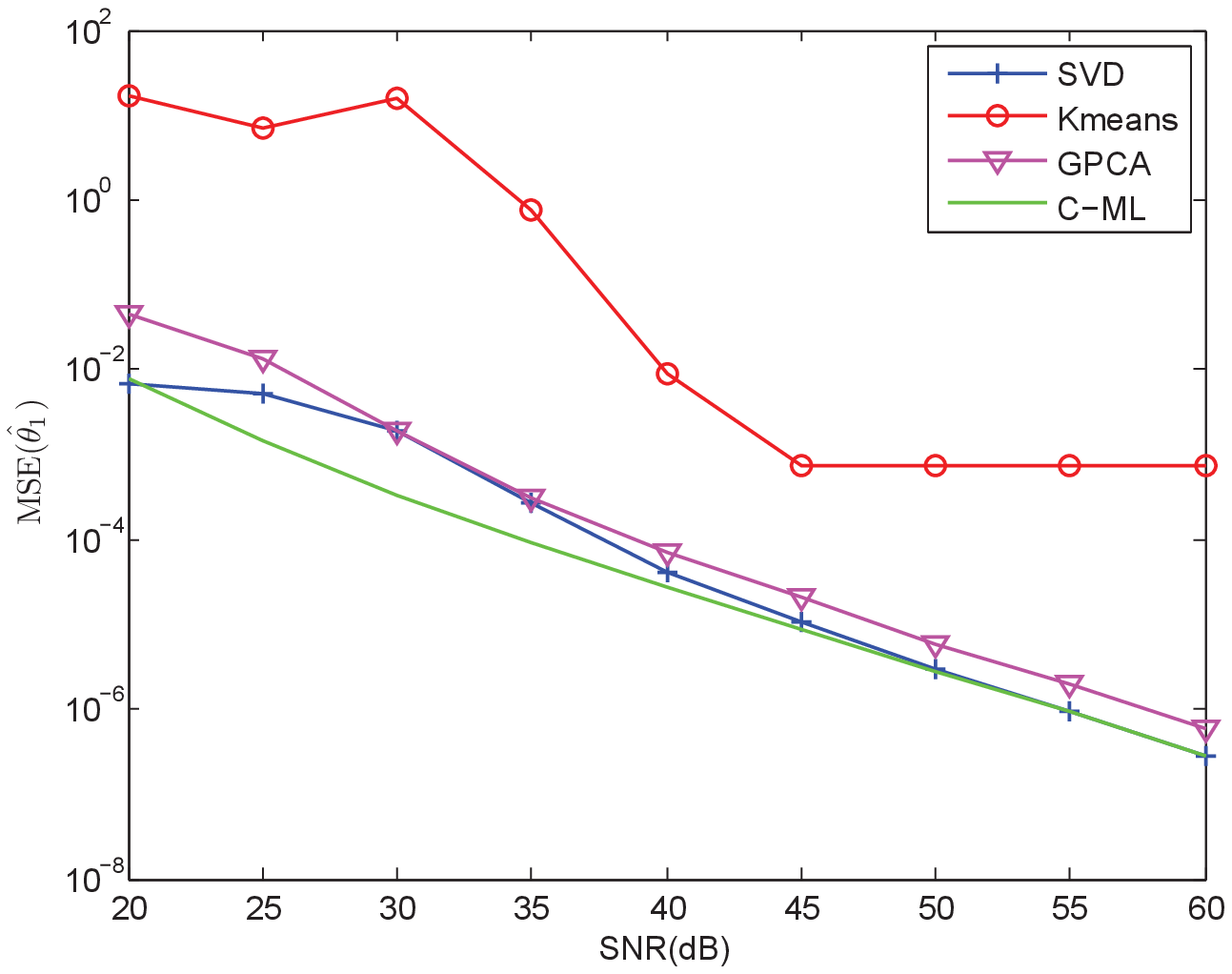}
  \includegraphics[width=0.48\hsize]{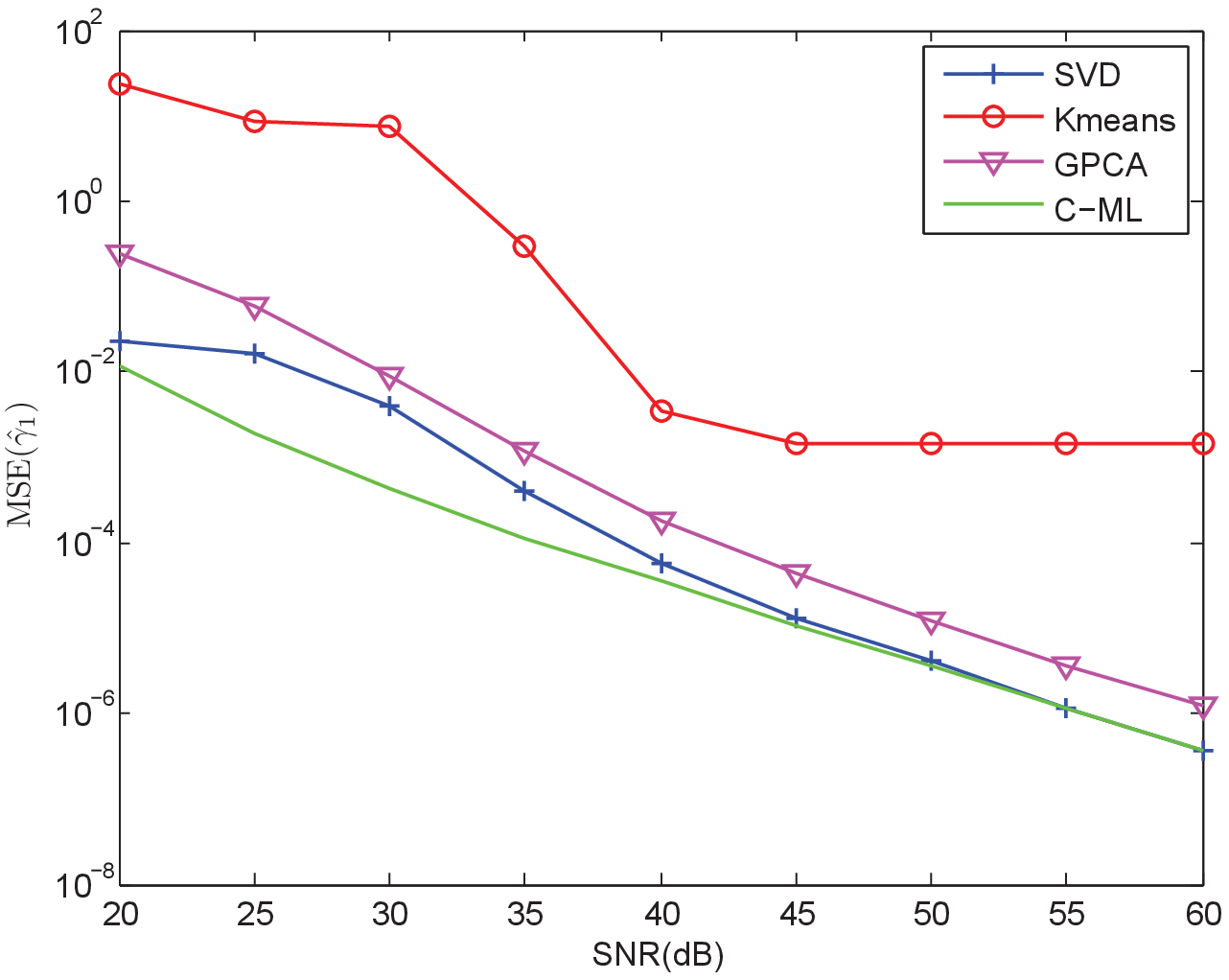}
  \includegraphics[width=0.48\hsize]{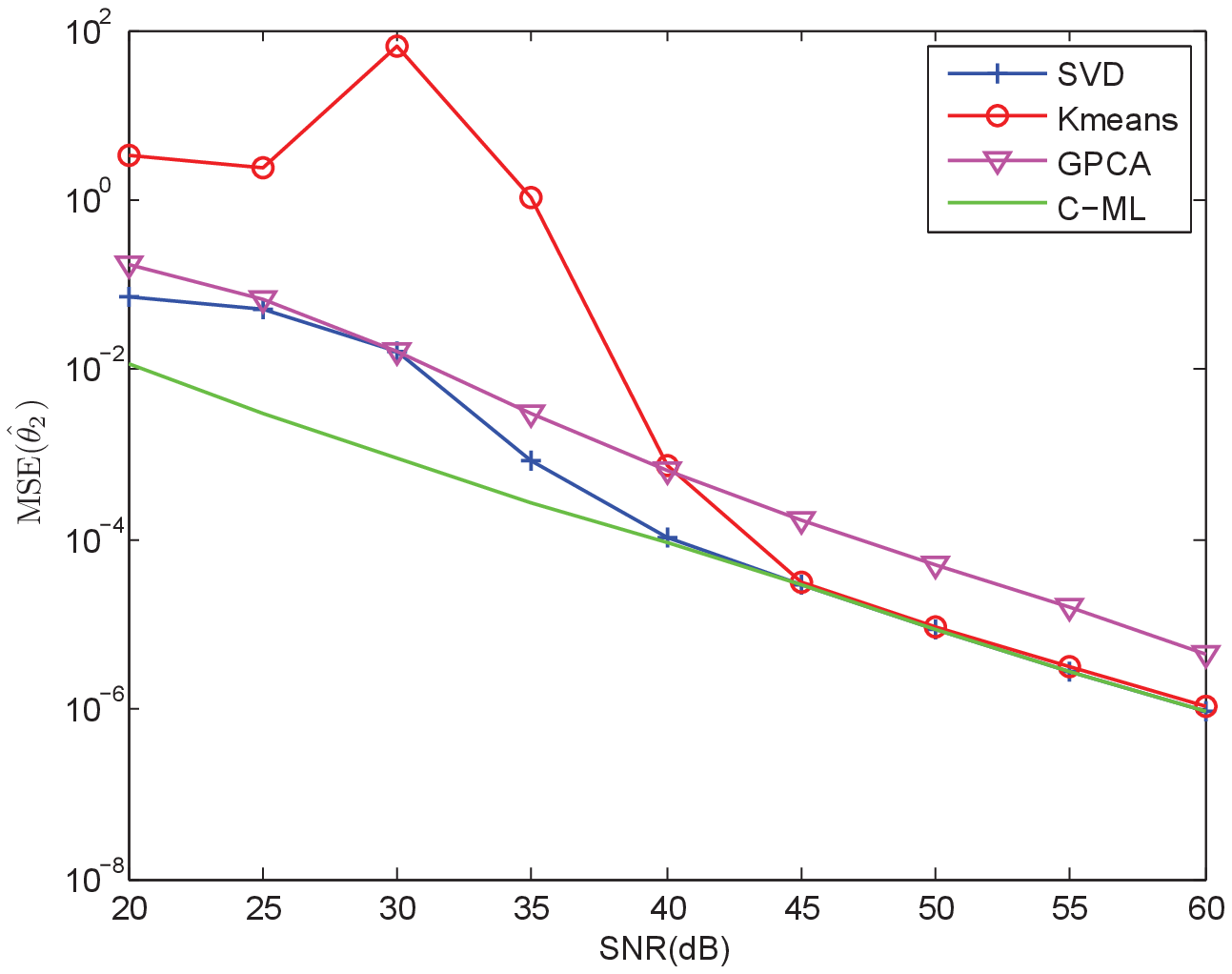}
  \includegraphics[width=0.48\hsize]{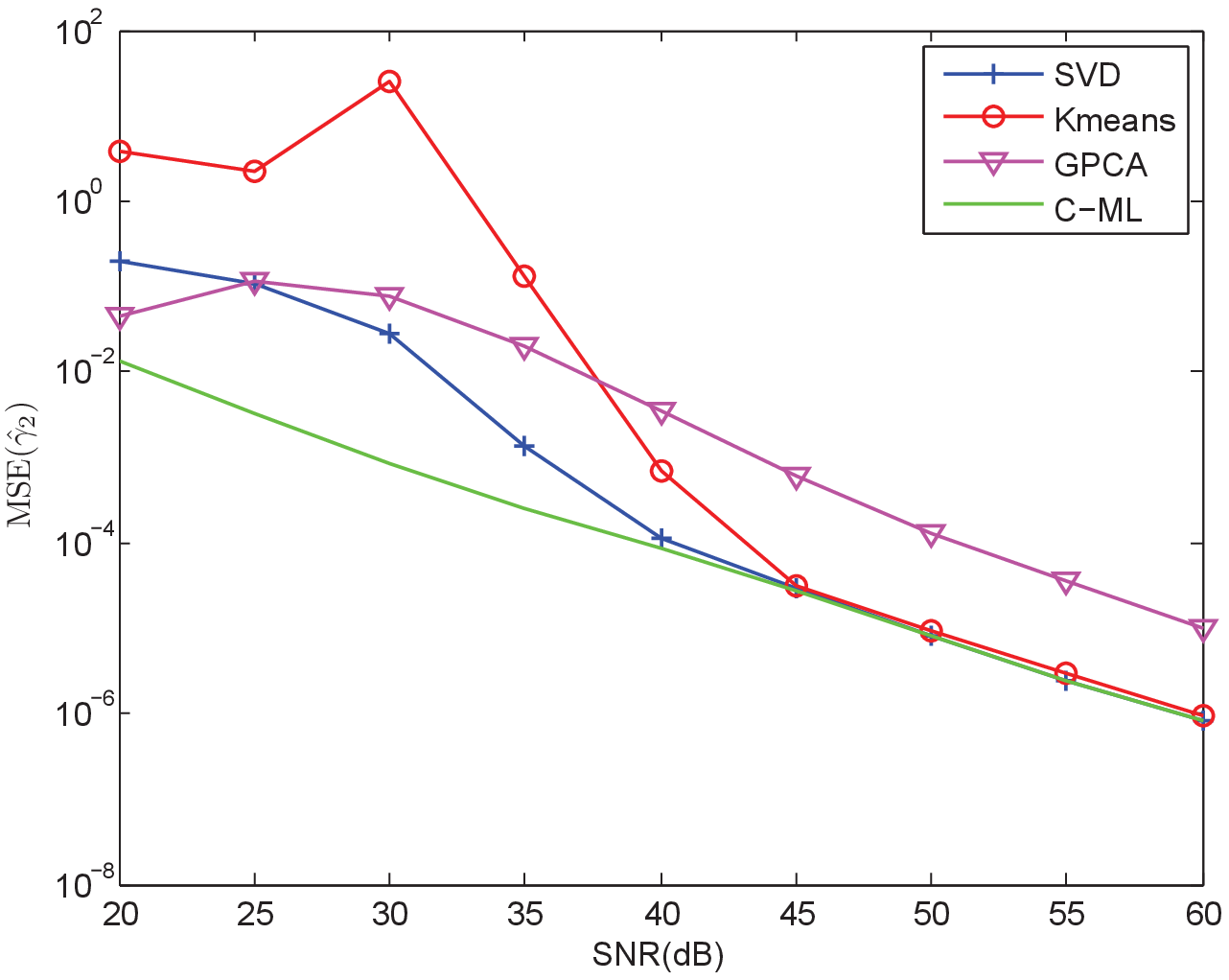}
  \label{fig_1D}
  \caption{\label{fig_1D_mse} Mean Squared Error (MSE) vs. SNR plots in Example 1. Sample amount is $N_1=100$, $N_2=100$, Monte Carlo runs is $10^4$. SCS: the spectral
  clustering on subspace algorithm; Kmeans: the $K$-mean based method in \citep{Ferrari}, the tuned coefficient of local data set was set as $c=7$; GPCA: generalized
  principal component analysis in \citep{Vidal2005}; C-ML: the maximum likelihood solution with the labels of observation known.}
\end{figure*}

Fig.~\ref{fig_1D_mse} shows the MSE plots of Algorithm.~\ref{alg1} against the benchmarks and  ``clairvoyant'' maximum likelihood estimation (C-ML).
We observed that the proposed algorithm achieved the C-ML at SNR greater than 35dB. We also know that, with Gaussian noise assumption, the C-ML achieves the Cram\'{e}r-Rao bound at high SNR. As expected, the subspace method had an SNR threshold below which its performance degrades, but still is better than benchmarks.  In this simulation, the threshold was about $35$dB.

The algebraic geometric method (GPCA) appeared to be slightly better at low SNR, but kept a constant gap to the C-ML estimation as SNR at very high SNR. The reasons for
this phenomena is that \textit{hybrid decoupling constraints} in  transforms the zero-mean noise into a nonzero-mean noise and cause a bias to the estimation results.
To clearly explain the MSE gap of algebraic geometric methods, we introduce the noise into the switched affine model in (\ref{eq:simple_a}) and (\ref{eq:simple_b}):
\begin{equation}
    S_1: \left\{\begin{array}{l}
                      x_n^{(1)} = d_n^{(1)}+e_n\\
                      y_n^{(1)}=\theta_1 d_n^{(1)} +\gamma_1+w_n
                    \end{array}
             \right.,
    S_2: \left\{\begin{array}{l}
              x_n^{(2)} = d_n^{(2)}+e_n \\
              y_n^{(2)}=\theta_2 d_n^{(2)} +\gamma_2+w_n
            \end{array}
     \right.
\end{equation}
and rewrite it in the form of \textit{hybrid decoupling constraints} (Since these constraints apply on all observations, the superscript is thrown away here) :
\begin{equation}
\left( {{y_n} - {\theta_1}\left( {{x_n} - {e_n}} \right) -\gamma_1- {w_n}} \right)\left( {{y_n} - {\theta_2}\left( {{x_n} - {e_n}} \right) -\gamma_2- {w_n}} \right) = 0
\end{equation}
\begin{align}
&\left(x_nw_n-x_ny_n-e_nw_n+e_ny_n\right)\left(\theta_1+\theta_2\right)\notag\\
&+\left(x_n-e_n\right)\left(\gamma_1\theta_2+\gamma_2\theta_1\right)-\left(y_n-w_n\right)\left(\gamma_1+\gamma_2\right)+\gamma_1\gamma_2\notag\\
&+\left(x_n^2-2x_ne_n+e_n^2\right)\theta_1\theta_2-\left(2y_nw_n-w_n^2\right)+y_n^2=0\notag\\
&\tilde \theta = \left[ {\begin{array}{c}
   {\theta_1\theta_2}  \\
   {\theta_1 + \theta_2}\\
   {\gamma_1\theta_2+\gamma_2\theta_1}\\
   {\gamma_1+\gamma_2}
 \end{array} } \right],\tilde \gamma=\gamma_1\gamma_2,\tilde x_n= \left[ {\begin{array}{c}
   {x_n^2}  \\
   {-x_ny_n} \\
   {x_n}\\
   {-y_n}
 \end{array} } \right],\tilde y_n = y_n^2\notag\\
&\tilde e_n = \left[ {\begin{array}{c}
   {2{x_n}{e_n} - e_n^2}  \\
   {-{x_n}{w_n} - {e_n}{y_n} + {e_n}{w_n}}\\
   {e_n}\\
   {-w_n}
\end{array} } \right], \tilde w_n = 2{y_n}{w_n} - {w_n}^2\notag
\end{align}
\begin{equation}
 {{\tilde y}_n} - \left( {{{\tilde x}_n} - {{\tilde e}_n}} \right)\tilde \theta -\tilde \gamma - {{\tilde w}_n} = 0
\end{equation}

We can find that the original noise $e_n$  and $w_n$ are squared in HDC operation. As new noise terms $E\{\tilde e_n\}\neq0$ and $E\{\tilde w_n\}\neq0$, algebraic
geometry based methods get a bias in this process. Besides, the original noises $e_n$  and $w_n$ are also amplified by the measurements in $\tilde e_n$ and $\tilde
w_n$. The Monte Carlo simulation in Fig.~\ref{fig_1D_mse} also verify the theoretical result above.

For the $K$-means based technique, the trend of decreasing MSE as the SNR increases halted at about SNR=45dB. As $K$-means based method relies on a local data
assumption, refer to \citep{Ferrari}, which transforms the observations into feature parameters. Because we set similar parameters in two submodels, outliers are
unavoidable for this technique even in noiseless case. From the MSE curves in Fig.~\ref{fig_1D_mse}, we can inference that some observations of the seconde submodel
were misclassified into the first submodel, whose performance was corrupted by these misclassified data. Although the loss of these data would cause a slight
degradation to the MSE of the second submodel, but it didn't worsen the estimation results.

\begin{figure}[thpb]
  \includegraphics[width=0.96\hsize]{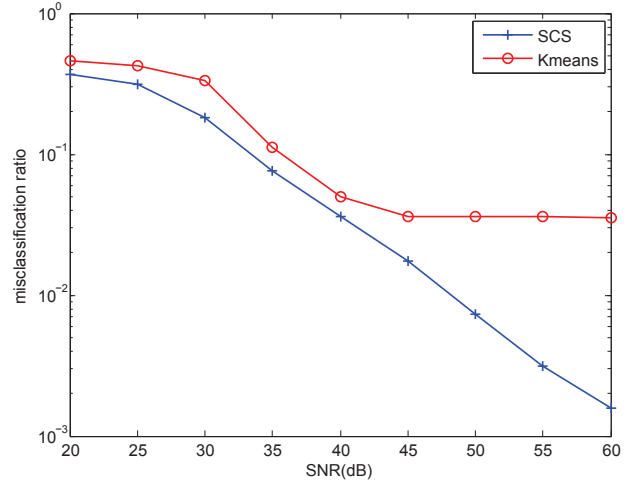}
  \caption{\label{fig_1D_mis} Misclassification ratio vs. SNR plots in Example 1.}
\end{figure}

We also can get a clear explanation from the misclassification ratio of both approaches, refer to Fig.~\ref{fig_1D_mis}. The $K$-mean based method in \citep{Ferrari}
kept a constant misclassification ratio when SNR is higher than 45dB, while SCS algorithm had misclassification error decays exponentially with respect to SNR.

\subsection{MIMO jump affine models with identical input domain}
A main advantage of the proposed subspace approach is that the input domain partition can be arbitrarily shaped
and it deals with MIMO system with relatively easily.  In this simulation, we consider an MIMO jump affine models with its submodels defined on the same domain. The
observation subspace of systems are given by
\begin{align}\label{case2}
  S_1:&\left\{\begin{array}{l}
    y_n=\left[\begin{array}{cc}
      0.7&0.4\\
      0.2&0.3
    \end{array}\right]d_n+\left[\begin{array}{c}
      -0.4\\
      0.17
    \end{array}\right]+w_n  \notag\\
    \notag\\
    x_n=\left[\begin{array}{cc}
      1&0\\
      0&1
    \end{array}\right]d_n+e_n
  \end{array}\right.  \notag\\
  S_2:&\left\{\begin{array}{l}
    y_n=\left[\begin{array}{cc}
      0.8&0.9\\
      0.4&0.5
    \end{array}\right]d_n+\left[\begin{array}{c}
      -0.81\\
      -0.09
    \end{array}\right]+w_n  \notag\\
    \notag\\
    x_n=\left[\begin{array}{cc}
      1&0\\
      0&1
    \end{array}\right]d_n+e_n
  \end{array}\right.
\end{align}

%
%\begin{figure}[hpb]
%  \centering
%  \includegraphics[width=0.8\hsize]{fig/check_board.eps}
%  \caption{\label{fig_chess}The chessboard domain in 2-D case. Sub-model one is defined on the shadow areas $\Phi_1$; sub-model two is defined on the plain areas
%  $\Phi_2$. }
%\end{figure}
\begin{figure*}[thpb]
  \centering
  \includegraphics[width=0.48\hsize]{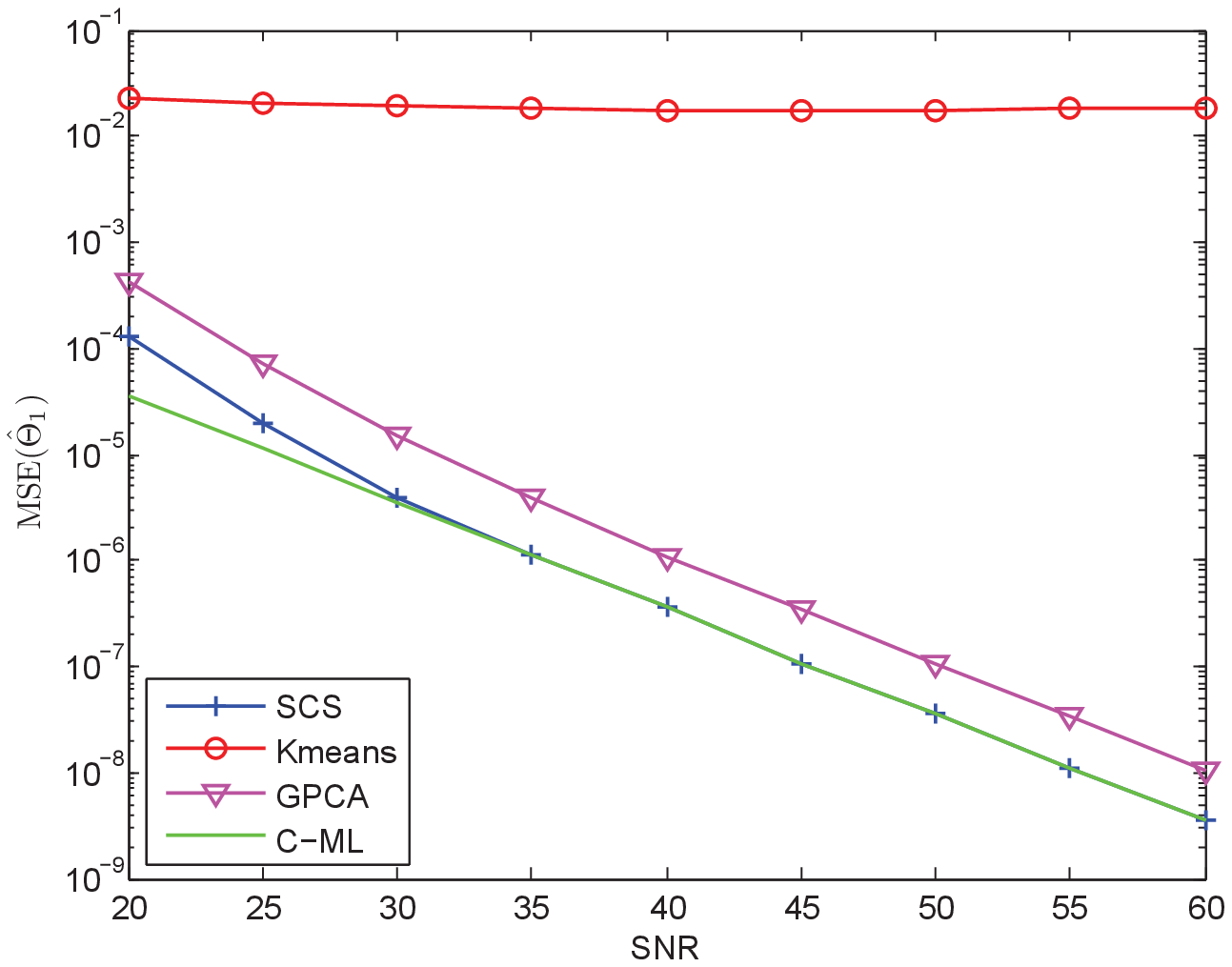}
  \includegraphics[width=0.48\hsize]{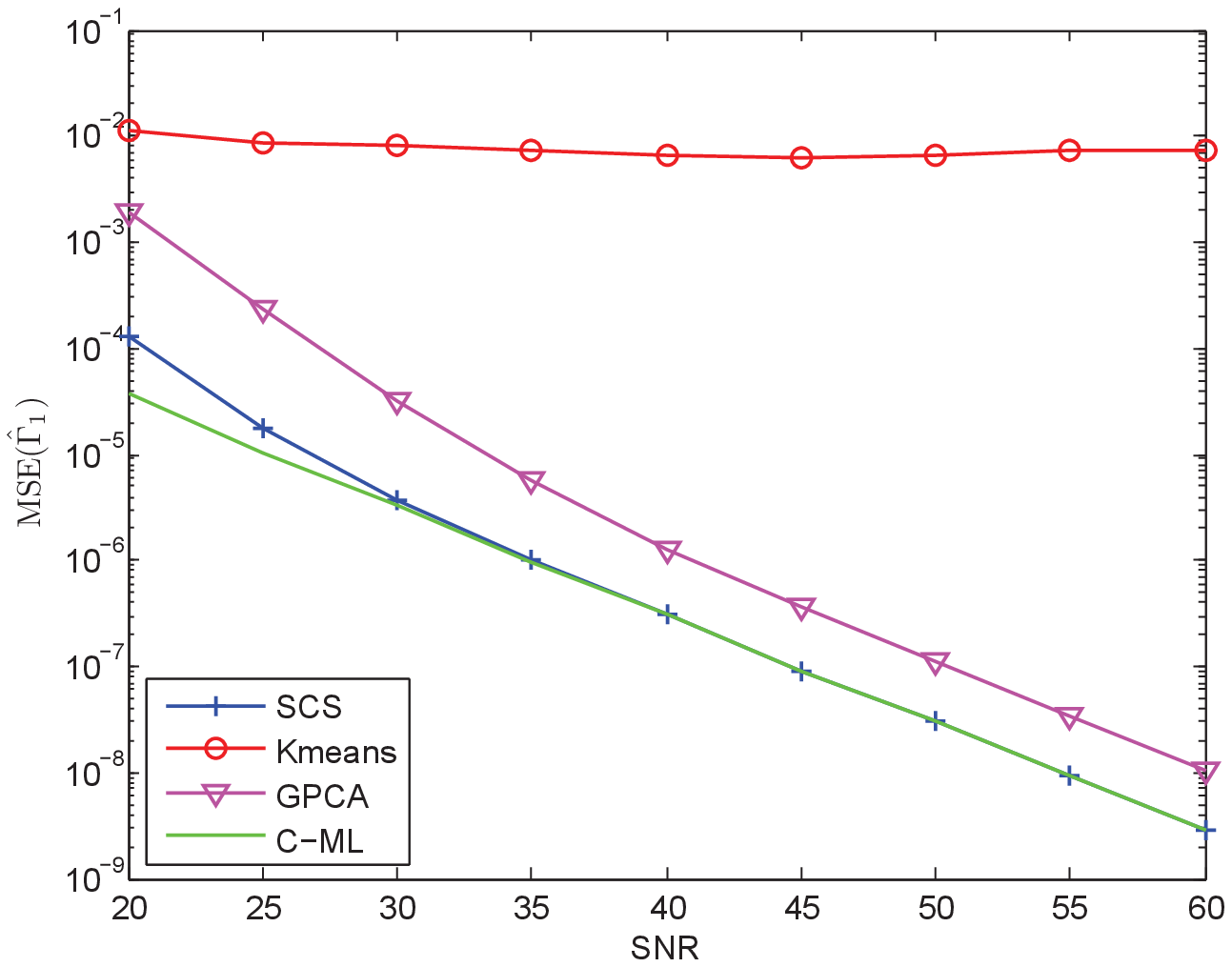}
  \includegraphics[width=0.48\hsize]{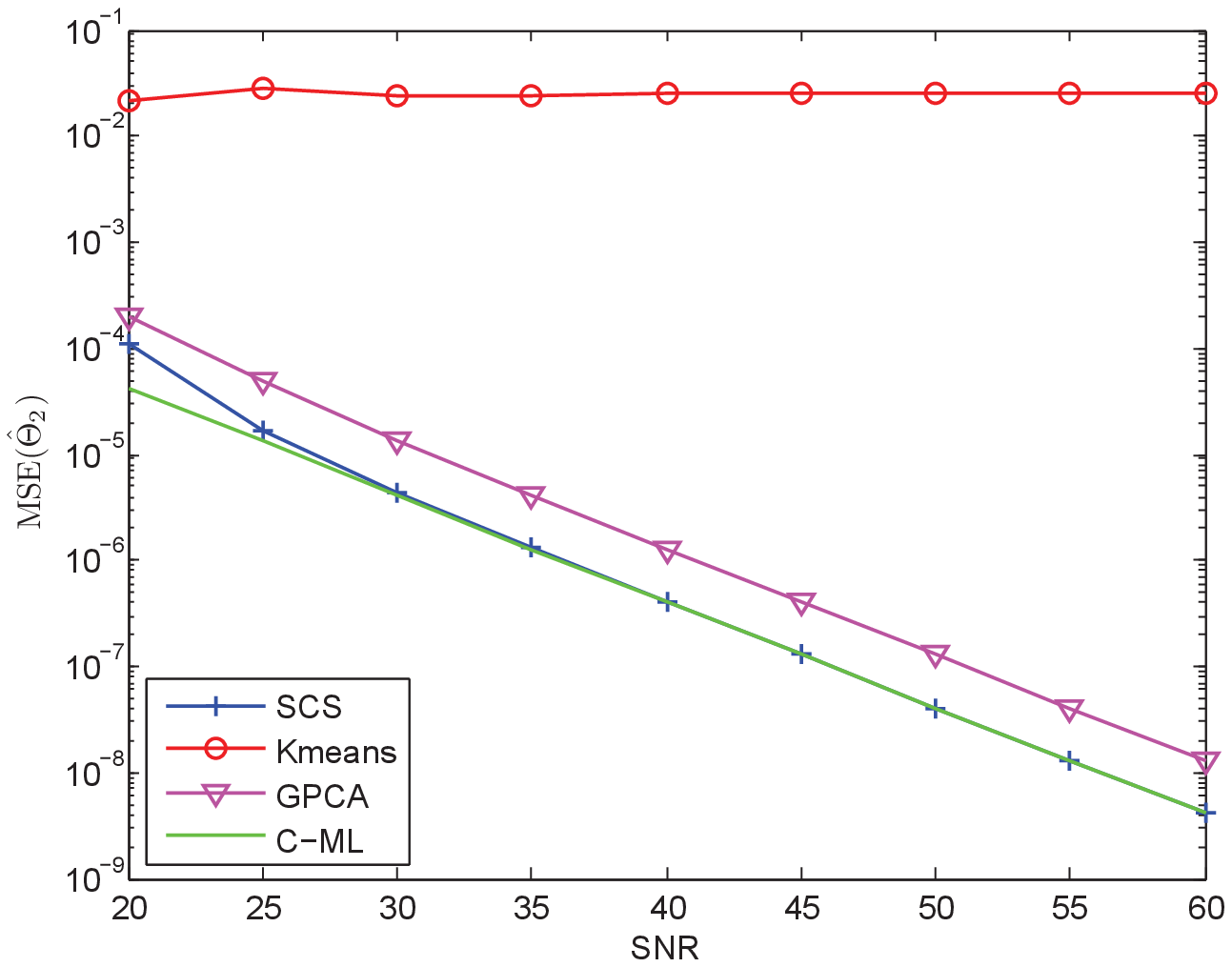}
  \includegraphics[width=0.48\hsize]{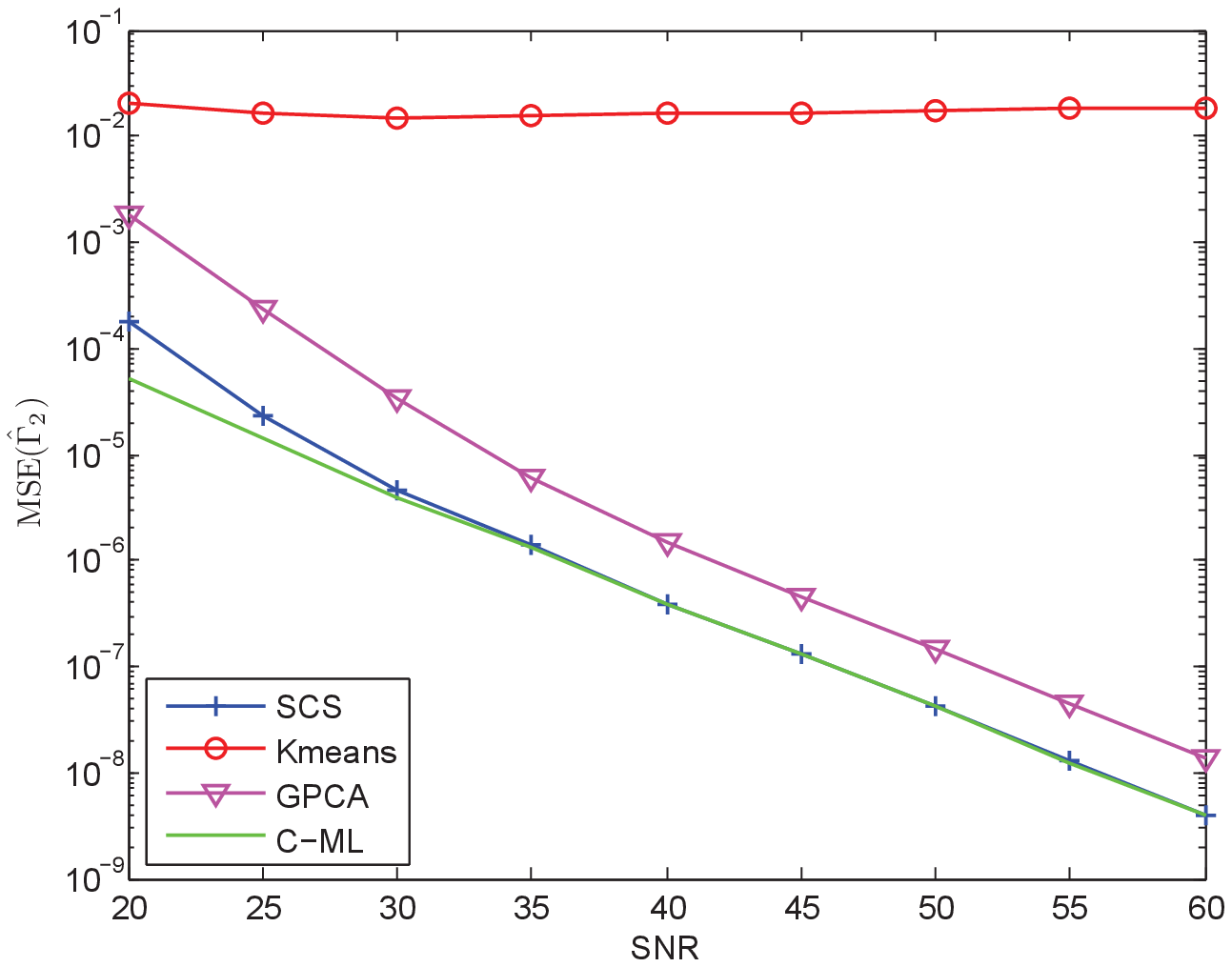}
  \caption{\label{fig_2D_mse}Estimated Average MSE of $\hat{\Theta}$ and $\hat{\Gamma}$ . Average MSE is the mean of all entries of parameter
  matrix. The data amount are $N_1=400$ and $N_2=400$, Monte Carlo runs is $10^3$ . SCS: the spectral clustering on subspace algorithm; Kmeans: the $K$-mean based
  method in \citep{Ferrari}, the local data set number is chosen as $c=10$; GPCA: generalized principal component analysis in \citep{Vidal2005}.
  C-ML: the maximum likelihood solution with the labels of observation known. The input sequences were generated by standard normal distribution.}
\end{figure*}

\begin{figure}[thpb]
  \centering
  \includegraphics[width=0.96\hsize]{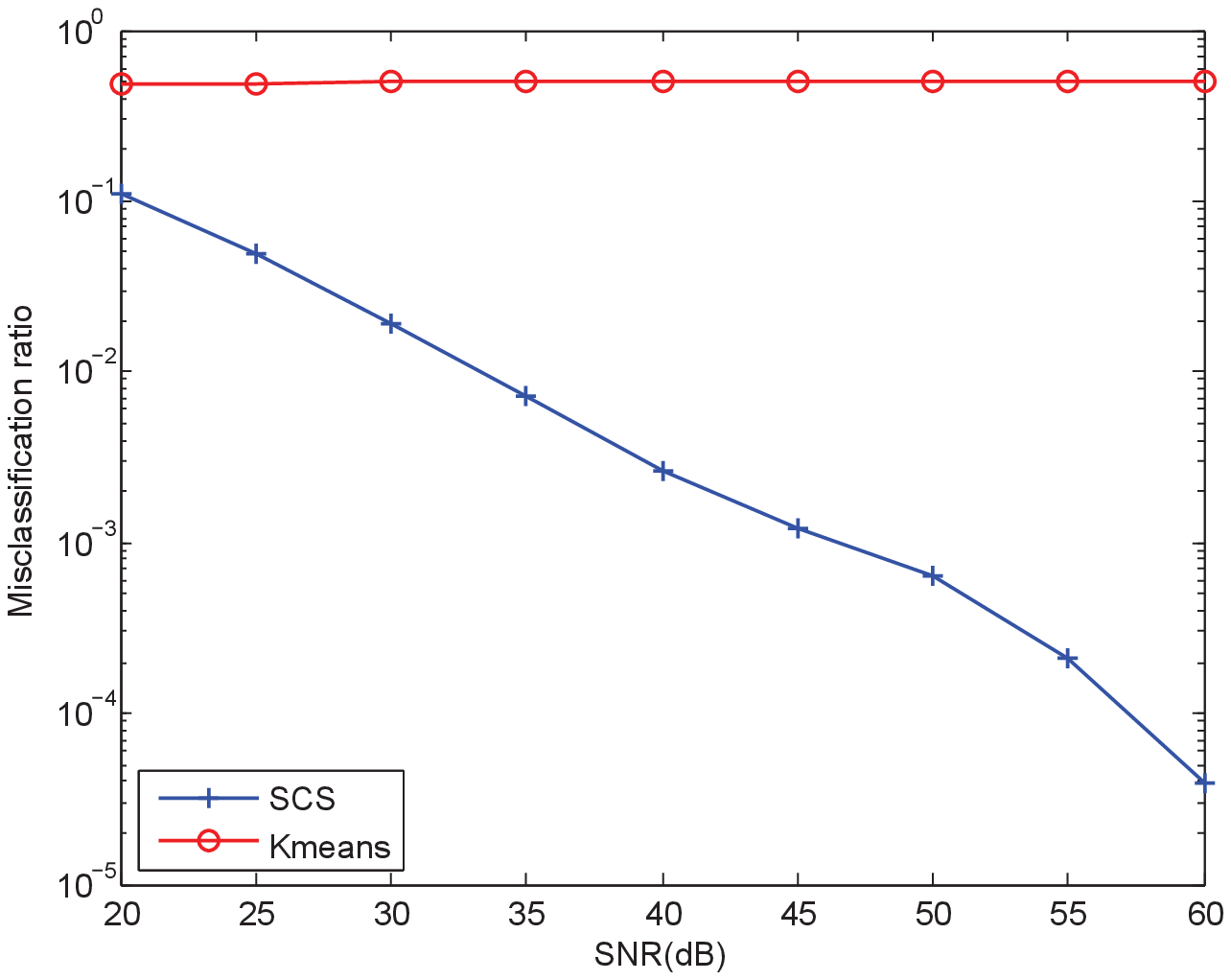}
  \caption{\label{fig_2D_mis}Misclassification ratio vs. SNR plots in Example 2.}
\end{figure}

The MSE performance comparisons are are shown in Fig.~\ref{fig_2D_mse}.  The proposed subspace estimator had  similar MSE performance as in the SISO case.  When the SNR
was greater than 30dB, the MSE performance matches with the clairvoyant C-ML.   $K$-means based technique didn't work in this case. Because observations of submodels
were mixed up on the whole domain, the local data assumption didn't hold in this scenario.  The MSE gap for Algebraic geometry techniques was further amplified because
the more parameters are involved in MIMO case. The misclassification ratio vs. SNR plot is shown in Fig.~\ref{fig_2D_mis}.  Misclassification ratio of $K$-means based
technique was kept around $50\%$, while proposed technique achieved a similar performance as in SISO case.

\section{Conclusions}

This paper extends the subspace clustering technique to parameter estimation of switched affine models. The first contribution of the proposed approach is the exploitation of a special subspace structure in the row space of the observation data matrix.  Then the process of associating data to sub-models is transformed into a matrix de-permutation problem.  The second contribution is investigating the subspace structure of the adjacency matrix $M=|VV^{\T}|$ and proposing spectral clustering on subspace algorithm to associate observations. The proposed algorithm guarantee identifiability in the absence of noise, and are shown to have improved performance against some existing benchmarks on handling the similar-parameter and arbitrary shaped domain cases. At last, the computation complexity of the proposed method is considerably reduced as the main operation is consisted of singular value decomposition.

The proposed algorithm applies only to the switched affine systems with sufficient number of independent output sensors and non-empty
intersection subspace.  This is a restriction not imposed by other methods. However, when these required conditions are satisfied, the proposed technique does have some computation and performance advantages.

\section*{Acknowledgment}

the National Key Technology R\&D Program under Grant 2009BAG12A08, the Research Foundation of the Ministry of Railways and Tsinghua University (RFMOR\&THU) under Grand 2009X003.

%%%%%%%%%%%%%%%%%%%%%%%%%%%%%%%%%%%%%%%%%%%%%%%%%%%%%%%%%%%%%%%%%%%%%%%%%%%%%%%%%%%%%%%%%
%%%%%%%%%%%%%%%%%%%%%%%%%%%%%%%%%%%%%%%%%%%%%%%%%%%%%%%%%%%%%%%%%%%%%%%%%%%%%%%%%%%%%%%%%
% references section

%%%%%%%%%%%%%%%%%%%%%%%%%%%%%%%%%%%%%%%%%%%%%%%%%%%%%%%%%%%%%%%%%%%%%%%%%%%%%%%%%%%%%%%%%
%%%%%%%%%%%%%%%%%%%%%%%%%%%%%%%%%%%%%%%%%%%%%%%%%%%%%%%%%%%%%%%%%%%%%%%%%%%%%%%%%%%%%%%%%

\section*{Appendix}
\subsection*{A. Proof of Theorem~1\label{apdex1}}
\begin{pf}
  The input signal space $\left(D-D_0\right)^{\T}P$ is related to the SVD of $Z$ by some full rank matrix $B, \left(D-D_0\right)^{\T}P=BV^{\T}$. We then have
  $\left(D-D_0\right)^{\T}\left(D-D_0\right)=BB^{\T}$. Because $D-D_0$ is block diagonal, so is $BB^T$, let
  \begin{equation}
    BB^{\T}=diag\left(\Psi_1^2, \ldots, \Psi_K^2\right)=\Psi^2, \Psi_i^2=\left(D_i-D_0\right)^{\T}\left(D_i-D_0\right).
  \end{equation}
  Normalize $B$ as
  \begin{equation}
    \bar{B}\triangleq \Psi^{-1}B, \bar{D}^{\T}\triangleq \Psi^{-1}\left(D-D_0\right)^{\T}
  \end{equation}
  We have
  \begin{equation}
    \bar{B}\bar{B}^{\T}=\bar{B}^{\T}\bar{B}=I
  \end{equation}
  From $\bar{D}^{\T}P=\bar{B}V^{\T}$, we have
  \begin{align}
    VV^{\T}&=P\bar{D}^{\T}\bar{D}P^{\T}\notag\\
    &=P\left[\begin{array}{ccc}
               \Lambda_1 &  & \\
               & \ddots & \\
               &  & \Lambda_K
             \end{array}
    \right]P^{\T}
  \end{align}
  where $\Lambda_i=\left(D_i-D_0\right)\left(\left(D_i-D_0\right)^{\T}\left(D_i-D_0\right)\right)^{-1}\left(D_i-D_0\right)^{\T}$
\end{pf}

\section*{References}
%\bibliographystyle{elsarticle-harv}
%\bibliographystyle{plain}
%\biboptions{number}
\bibliographystyle{elsarticle-num}
\biboptions{sort&compress}
\bibliography{ref}
\end{document}